\documentstyle[preprint,eqsecnum,tighten,aps]{revtex}
\def\MEts{{\mbox{$E\kern-0.57em\raise0.19ex\hbox{/}_{T}$}}}
\def\MEt{{\mbox{$E\kern-0.57em\raise0.19ex\hbox{/}_{T}$}}\ }
\def\MEtc{{\mbox{$E\kern-0.57em\raise0.19ex\hbox{/}_{T}^{\it cal}$}}\ }
\def\MEtg{{\mbox{$E\kern-0.57em\raise0.19ex\hbox{/}_{T}\gamma$}}\ }

\begin{document}
\draft
\title{Study of the $ZZ\gamma$ and $Z\gamma\gamma$ Couplings in $Z(\nu\nu)\gamma$
Production}
\preprint{Fermilab--Pub--97/047--E}

%
\author{                                                                        
S.~Abachi,$^{14}$                                                               
B.~Abbott,$^{28}$                                                               
M.~Abolins,$^{25}$                                                              
B.S.~Acharya,$^{43}$                                                            
I.~Adam,$^{12}$                                                                 
D.L.~Adams,$^{37}$                                                              
M.~Adams,$^{17}$                                                                
S.~Ahn,$^{14}$                                                                  
H.~Aihara,$^{22}$                                                               
G.A.~Alves,$^{10}$
% ...
\\                                                                              
\vskip 0.50cm                                                                   
\centerline{(D\O\ Collaboration)}                                               
\vskip 0.50cm                                                                   
}                                                                               
\address{                                                                       
\centerline{$^{1}$Universidad de los Andes, Bogot\'{a}, Colombia}               
\centerline{$^{2}$University of Arizona, Tucson, Arizona 85721}                 
\centerline{$^{3}$Boston University, Boston, Massachusetts 02215}               
\centerline{$^{4}$Brookhaven National Laboratory, Upton, New York 11973}        
\centerline{$^{5}$Brown University, Providence, Rhode Island 02912}             
\centerline{$^{6}$Universidad de Buenos Aires, Buenos Aires, Argentina}         
\centerline{$^{7}$University of California, Davis, California 95616}            
\centerline{$^{8}$University of California, Irvine, California 92717}           
\centerline{$^{9}$University of California, Riverside, California 92521}        
\centerline{$^{10}$LAFEX, Centro Brasileiro de Pesquisas F{\'\i}sicas,          
                  Rio de Janeiro, Brazil}                                       
\centerline{$^{11}$CINVESTAV, Mexico City, Mexico}                              
\centerline{$^{12}$Columbia University, New York, New York 10027}               
\centerline{$^{13}$Delhi University, Delhi, India 110007}                       
\centerline{$^{14}$Fermi National Accelerator Laboratory, Batavia,              
                   Illinois 60510}                                              
\centerline{$^{15}$Florida State University, Tallahassee, Florida 32306}        
\centerline{$^{16}$University of Hawaii, Honolulu, Hawaii 96822}                
\centerline{$^{17}$University of Illinois at Chicago, Chicago, Illinois 60607}  
\centerline{$^{18}$Indiana University, Bloomington, Indiana 47405}              
\centerline{$^{19}$Iowa State University, Ames, Iowa 50011}                     
\centerline{$^{20}$Korea University, Seoul, Korea}                              
\centerline{$^{21}$Kyungsung University, Pusan, Korea}                          
\centerline{$^{22}$Lawrence Berkeley National Laboratory and University of      
                   California, Berkeley, California 94720}                      
\centerline{$^{23}$University of Maryland, College Park, Maryland 20742}        
\centerline{$^{24}$University of Michigan, Ann Arbor, Michigan 48109}           
\centerline{$^{25}$Michigan State University, East Lansing, Michigan 48824}     
\centerline{$^{26}$Moscow State University, Moscow, Russia}                     
\centerline{$^{27}$University of Nebraska, Lincoln, Nebraska 68588}             
\centerline{$^{28}$New York University, New York, New York 10003}               
\centerline{$^{29}$Northeastern University, Boston, Massachusetts 02115}        
\centerline{$^{30}$Northern Illinois University, DeKalb, Illinois 60115}        
\centerline{$^{31}$Northwestern University, Evanston, Illinois 60208}           
\centerline{$^{32}$University of Notre Dame, Notre Dame, Indiana 46556}         
\centerline{$^{33}$University of Oklahoma, Norman, Oklahoma 73019}              
\centerline{$^{34}$University of Panjab, Chandigarh 16-00-14, India}            
\centerline{$^{35}$Institute for High Energy Physics, 142-284 Protvino, Russia} 
\centerline{$^{36}$Purdue University, West Lafayette, Indiana 47907}            
\centerline{$^{37}$Rice University, Houston, Texas 77005}                       
\centerline{$^{38}$Universidade Estadual do Rio de Janeiro, Brazil}             
\centerline{$^{39}$University of Rochester, Rochester, New York 14627}          
\centerline{$^{40}$CEA, DAPNIA/Service de Physique des Particules, CE-SACLAY,   
                   France}                                                      
\centerline{$^{41}$Seoul National University, Seoul, Korea}                     
\centerline{$^{42}$State University of New York, Stony Brook, New York 11794}   
\centerline{$^{43}$Tata Institute of Fundamental Research,                      
                   Colaba, Mumbai 400005, India}                                
\centerline{$^{44}$University of Texas, Arlington, Texas 76019}                 
\centerline{$^{45}$Texas A\&M University, College Station, Texas 77843}         
}                                                                               

\maketitle
\vspace{-0.2in}
\begin{abstract}     
We have measured the  $ZZ\gamma$ and  $Z\gamma\gamma$ couplings by studying
$p\bar p \to  \MEts\gamma + X$  events at  $\sqrt{s}=1.8$~TeV with the D\O\
detector at the  Fermilab Tevatron  Collider. This  first study of hadronic
$Z\gamma$ production in the neutrino decay channel gives the most stringent
limits on  anomalous  couplings available. A  fit to the  transverse energy
spectrum  of  the  photon in  the candidate event sample,  based  on a  data set
corresponding to an  integrated luminosity of  $13.1\ {\rm pb}^{-1}$, yields
$95\%$ CL limits  on the anomalous ${\it CP}$-conserving
$ZZ\gamma$    couplings of    $|h^Z_{30}|<0.9$,    $|h^Z_{40}|<0.21$, for a
form-factor  scale $\Lambda =  500$~GeV.  Combining these  results with our
previous   measurement using $Z  \to ee$  and  $\mu\mu$ yields  the limits:
$|h^Z_{30}|<0.8$,         $|h^Z_{40}|<0.19$    ($\Lambda =    500$~GeV) and
$|h^Z_{30}|<0.4$, $|h^Z_{40}|<0.06$ ($\Lambda = 750$~GeV). 
\end{abstract}         
\pacs{\it Submitted to Phys. Rev. Lett.}

In the Standard Model (SM),  couplings of the form $ZV\gamma$, where $V$ is
a $Z$  or  $\gamma$,  vanish  at  tree  level.  Direct   measurement of the
$ZV\gamma$  couplings
is  made possible by
studying  $Z\gamma$ production.  Previously, only the  charged lepton decay
modes of the $Z$ have been studied  in $p\bar p$ collisions at the Tevatron
($\sqrt{s}=1.8$~TeV)~\cite{CDF-Zg,D0-Zg}.   
Here we report 
the first measurement of  $Z\gamma$ production in  the invisible (neutrino)
decay channel of the $Z$ at a hadron collider;
such studies have recently been made at
LEP~\cite{L3-Zg,DELPHI-Zg}. This
analysis of the  neutrino decay  channel significantly improves
the  limits  on   $ZZ\gamma$ and    $Z\gamma\gamma$   trilinear  couplings
and, in combination with previous D\O\  limits from other decay
channels~\cite{D0-Zg}, gives stringent new limits.

We have studied   the reaction 
$p\bar p \to
\MEts\gamma + X$ (where \MEt is  missing transverse energy) using data from
the 1992--1993 Tevatron run with
the D\O\ detector, corresponding to
an exposure of $13.1 \pm 0.7~{\rm pb}^{-1}$. The advantages of
using the $Z \to \nu \nu$ mode compared with the $\ell^+ \ell^-$
decay channels are larger geometrical acceptance and detection efficiency;
higher  branching ratio  (by a  factor of six  over $ee$ or  $\mu\mu$); and
absence of the  radiative  $Z$-decay  contribution.  However, the invisible
decay mode of the $Z$ does not allow reconstruction of the $Z$ mass and has
larger potential background.

The D\O\ detector,  described in detail  elsewhere~\cite{ref3}, consists of
three main systems. Central and forward drift chambers are used to identify
charged tracks  for $|\eta|\leq  3.2,$ where $\eta$ is  pseudorapidity. The
calorimeter consists of  uranium-liquid argon  sampling detectors with fine
segmentation    in a  central  and  two  end   cryostats,  and  provides
near-hermetic  coverage for  $|\eta|\leq 4.4.$ The  energy resolution of the
calorimeter was   measured in   beam     tests~\cite{ref4}  to be
$15\%/\sqrt{E}$ for electrons and $50\%/\sqrt{E}$ for isolated pions
($E$ in GeV). The calorimeter towers subtend $0.1\times
0.1$ in $\eta \times \phi$ ($\phi$ is the azimuthal angle), segmented
longitudinally  into four  electromagnetic  (EM) and four  or five hadronic
layers. In  the third EM  layer, at  the EM  shower maximum,  the cells are
$0.05\times   0.05$  in   $\eta\times\phi$.  The  muon  system  consists of
magnetized iron toroids with one inner and two outer layers of drift tubes,
providing   coverage for   $|\eta|\leq  3.3.$ For  this  analysis  the muon
detector was used only as a veto.

$Z\gamma$  candidates were  selected by  requiring a  significant amount of
\MEt and an  isolated photon  with high  transverse energy  ($E_T^\gamma$).
There are three major sources of background to $\MEts\gamma$ production: 1)
jet- ($j$) related background from  $jj$ and $j\gamma$ production, occurring
when a jet hits a poorly  instrumented region of  the detector resulting in
mismeasured   \MEts. In the  dijet  case, one  jet  additionally  has to be
reconstructed as a photon when fragmenting into a leading neutral meson; 2)
cosmic ray or  beam halo muon   bremsstrahlung in the EM  calorimeter which
results  in a   reconstructed single  photon  in the  event with  balancing
missing  energy; 3) $W$  boson  production (with  $W \to  e\nu$), where the
electron is  reconstructed as a photon due to  inefficiency of the tracking
chambers.   Other  backgrounds,  such as   $W(\mu\nu)+j$  or  $Z(\nu\nu)+j$
production with a  jet faking a  photon (and an  unreconstructed or forward
muon for the $W$ case) are negligible.

The  $\MEts\gamma$  sample was  obtained with a  trigger  which required an
isolated EM cluster with  $E_T \geq 20$~GeV. A  photon cluster was required
to be within the  fiducial region of the  calorimeter and tracking chambers
($|\eta|\leq 1.0$  in the central  calorimeter (CC) or  $1.5\leq |\eta|\leq
2.5$ in  the end  calorimeters  (EC)).  The offline  photon  identification
requirements  were:  ({\it i\/})  EM energy  $> 0.96$ times
the total shower  energy;  ({\it  ii\/})  lateral and  longitudinal
shower shape consistent with  that of an electron  shower~\cite{ref3};
({\it iii\/})  the isolation  variable of  the  cluster~\cite{D0-Zg} $<
0.1$; ({\it iv\/}) a photon  cluster with no evidence of associated tracks
or hits in the drift chambers; ({\it v\/}) development of the photon shower
in the EM  calorimeter consistent with its origin  at the interaction
vertex reconstructed by the tracking  chambers; ({\it vi\/}) no muon tracks
in the central  calorimeter near the photon; ({\it vii\/}) no
additional EM  clusters in the event with $E_T > 5$~GeV; and ({\it viii\/})
$E_T^\gamma > 40$~GeV.

Missing   energy was   calculated  using  the    ca\-lo\-ri\-me\-ter energy
deposits. The  hadronic calorimeter  energy scale 
was  determined by  minimizing  the  average \MEt in  inclusive  $Z \to ee$
events. The resolution  of the  missing transverse energy projected
on a given axis was  $\approx 6$~GeV and depended  slightly on the boost of
the  $Z\gamma$ system.  
We required \MEt to exceed
40~GeV. We also  required no  reconstructed muons in  the central region of
the detector  ($|\eta_\mu| < 1.0$)  and no additional  hadronic jets in the
event with transverse energies above 15~GeV.

This selection resulted in  four  $Z(\nu\nu)\gamma$ candidates. Three
events had a photon in the CC and one in the EC. The highest photon $E_T$ in
this sample was 68 GeV.

To estimate the  number of  surviving  jet-related  background events,
we first determined the probability to mismeasure
\MEt in the  detector
by comparing the numbers  of $\MEts j$ and
$jj$ events collected in the same data set.  This probability
falls exponentially with \MEt and is  $< 10^{-4}$ for $\MEts > 35$~GeV. The
probability for a jet to fake a  photon was measured~\cite{D0-Zg,Greg} to be
$(7 \pm 2)\times 10^{-4}$. These  probabilities were applied to the $j\gamma
+ X$ cross section~\cite{Owens} and  $jj + X$ cross section (calculated from
data) with a  minimum transverse  energy cut of 40 GeV  imposed on jets and
photons. The total background from these sources was estimated to be $<0.6$
events.

The muon   bremsstrahlung background  was  significantly  suppressed by the
photon quality  criteria  ({\it v\/}) and  ({\it vi\/}), as  well as by the
high $E_T^\gamma$ cut and the  central muon veto.  (The muon veto was
not applied in the forward region  due to
high chamber occupancy.)
Muon bremsstrahlung backgrounds were
reduced by requiring that the photon direction deduced from the finely
divided EM calorimeter be consistent with the event vertex location.
The photon impact parameter resolution was  10--20~cm.
Additional  suppression  of the  cosmic ray
background  was   achieved by  rejecting  events  with a   muon-like energy
deposition in the  vicinity of the photon  cluster. The residual background
was estimated by applying the photon  quality cuts to very clean samples of
muon  bremsstrahlung events. The  estimated total muon  background is $1.8
\pm 0.6$ events.

The $W \to  e\nu$ background  was  suppressed by the  $E_T^\gamma$ and \MEt
cuts, set above the  Jacobian peak for $W \to  e\nu$ decays, and by the jet
veto which  decreased the smearing  of the Jacobian  peak due to associated
jet  production. It was  further  reduced by the  photon  quality cut ({\it
iv\/})  which  rejected  photons  with   associated  tracks or  hits in the
tracking chambers within  roads pointing to the  EM cluster. The  rejection
power of these cuts was estimated using $Z \to ee$ and $W \to e\nu$ samples
with electrons reconstructed as  photons due to the absence of a track. The
residual  background was estimated  using the $W \to  e\nu$ sample with the
cuts similar to the ones used for signal (except that a reconstructed track
was required  to match the EM  cluster). The  number of  background events,
obtained by  applying  track- and  hit-counting  rejection  factors to this
sample, was estimated to be $4.0 \pm 0.8$ events. 

The total muon and $W \to e\nu$  background is $5.8 \pm 1.0$ events. Since
the total  jet-related background  was less than the  error on the dominant
backgrounds, it was (conservatively)  neglected when deriving the limits on
the couplings. Table~\ref{table1} summarizes the backgrounds.

The acceptance of the D\O\ detector  for the $\nu\nu\gamma$ final state was
determined using the leading order  event generator \cite{ref2} to generate
4-vectors  for  the  $Z\gamma$   processes as a  function  of the  coupling
parameters. The 4-vectors were used  as input to a fast detector simulation
program which modeled  the effects of the EM and  missing transverse energy
resolutions,   interaction  vertex  spread, and  offline  efficiencies. The
efficiencies were estimated primarily by using $Z\to ee$ data. The trigger was
fully efficient  for $E_T^\gamma >  40$~GeV. The  overall efficiency of the
photon selection cuts was $0.57 \pm 0.03$ ($0.64 \pm 0.05$) in CC (EC). The
geometrical acceptance was 80\% for  the SM case and increased slightly for
non-zero    couplings.  The  ${\rm     MRSD-}'$~\cite{MRSD}  set of  parton
distribution functions (pdf) was  used in the calculations. The uncertainty
due to the choice of pdf  (6\%, determined by  using different pdf choices)
was included  in the  systematic error of  the Monte Carlo  calculation. We
accounted for the effect of higher order QCD corrections by multiplying the
rates by a constant factor $k =  1.34$~\cite{ref2}. The jet veto efficiency was
estimated to  be $0.84 \pm  0.02$ by  applying the veto  requirement to the
inclusive $Z \to ee$  data. The value of the  $k$-factor and the efficiency
of the jet veto  were shown to be  consistent with the  NLL $Z\gamma$ Monte
Carlo~\cite{Ohnemus} for the SM couplings.

The expected signal for SM couplings is $1.8 \pm 0.2 \pm 0.1$ events, where
the first  error is  due to the   uncertainty in the  Monte  Carlo modeling
(13\%),  and  the  second   is  the   uncertainty in  the  integrated
luminosity   calculation  (5.4\%). Our  observed  signal agrees  within the
errors with  the background  expectation  plus the SM  prediction.
We verified this by  simultaneously modifying the  cuts on $E_T^\gamma$
and \MEt to  35~GeV or 45~GeV; in  both cases the  observed number of events
agreed  well with  the  predictions. The  $E_T$  spectrum of  the candidate
events along with  the SM prediction  and estimated  background is shown in
Fig.~\ref{fig1}.

The most general Lorentz and gauge invariant $ZV\gamma$ vertex is described
by four coupling parameters  $h^V_i$~\cite{ref1}. Combinations of the ${\it
CP}$-conserving   (${\it  CP}$-violating)   parameters $h^V_3$  and $h^V_4$
($h^V_1$  and $h^V_2$)  correspond  to the  electric  (magnetic) dipole and
magnetic (electric) quadrupole transition moments of the $ZV\gamma$ vertex.
Non-zero (anomalous) values of the  $h^V_i$ couplings result in an
increase of the $Z\gamma$ production  cross section, particularly for large
$E_T^\gamma$~\cite{ref2}.  Partial wave unitarity  of the general $f{\bar
f} \to  Z\gamma$ process  restricts  the  $ZV\gamma$  couplings uniquely to
their vanishing SM values at asymptotically high energies~\cite{unitarity}.
Therefore, the coupling parameters  must be modified by form-factors $h^V_i
= h^V_{i0} / (1 +  \hat{s}/\Lambda^2)^n$, where  $\hat{s}$ is the square of
the invariant  mass of the $Z\gamma$  system, $\Lambda$  is the form-factor
scale,   and     $h^V_{i0}$  are   coupling    values at   the  low  energy
limit~\cite{ref2}.    
We take $n  = 3$ for $h^V_{1,3}$  and $n = 4$  for  $h^V_{2,4}$~\cite{ref2}.  
This  choice  yields the same
asymptotic  energy  behavior  for all of  the  couplings.  Unlike $W\gamma$
production where  form-factor  effects do not play  a crucial role, the
$\Lambda$-dependent  effects cannot  be ignored in  $Z\gamma$ production at
Tevatron  energies. This  is due to  the higher  power of  $\hat{s}$ in the
vertex  function, a  direct  consequence  of the  additional  Bose-Einstein
symmetry of the $ZV\gamma$ vertices~\cite{ref2}.

To set limits on  the anomalous couplings, a  fit to the observed $E_T$
spectrum of  the photon  with the  Monte Carlo  signal  prediction plus 
estimated   background  was  done. The  fit was   performed using  a binned
likelihood  method~\cite{Greg}, with Poisson  statistics for the signal and
Gaussian uncertainties for background, luminosity and efficiencies. Because
the  contribution of the  anomalous  couplings is  concentrated in the high
$E_T^\gamma$ region, the differential distribution $d\sigma/dE_T^\gamma$ is
more sensitive to the anomalous couplings than the total cross section (see
inset in  Fig.~\ref{fig1} and  Ref.~\cite{ref2}). To  exploit the fact that
anomalous  coupling contributions  lead to an excess of  events with a high
$E_T$  photon,  a   high-$E_T^\gamma$  bin, with  no  events  observed, was
explicitly included in the fit~\cite{Greg}. 

The  one- and  two-degree  of  freedom (DOF)  95\% CL  limits on  anomalous
couplings in the  $(h^Z_{30},h^Z_{40})$ plane  were obtained by cutting the
likelihood  function 1.92 or  3.00 units  below the  maximum. A form-factor
scale of $\Lambda  = 500$~GeV was  used in these  calculations. The two-DOF
limit  contour (see   Fig.~\ref{fig2}a)  represents the {\it  correlated\/}
limit on a pair of  couplings when both are  allowed to vary independently.
For models which predict  a particular  relationship between the couplings,
thus  eliminating  one DOF,  the  appropriate  point on  the  one-DOF limit
contour should be used. The limit on one coupling when all others are fixed
at the SM  values is  given by the  intersection  of this  contour with the
corresponding  axis ({\it axis\/}  limit). Since the  $(h^Z_{30},h^Z_{40})$
pair  is  nearly   uncorrelated  with  the  other    pairs~\cite{D0-Zg} the
correlated limits in the  above plane are a good  approximation of the {\it
global\/} limits, i.e. limits independent of the values of other couplings.
In what follows only  axis limits are quoted; the  correlated limits can be
obtained   from  the  figures.  The  95\%  CL  axis  limits  for  the ${\it
CP}$-conserving    $ZZ\gamma$  and    $Z\gamma\gamma$  couplings  from this
measurement    are   listed in      Table~\ref{table2}.  Limits  on  a {\it
CP\/}-violating   pair of  couplings are   numerically the same  as for the
corresponding {\it CP\/}-conserving pair.

Combined  limits on  anomalous couplings  were also  obtained based on this
measurement and previous D\O\ results~\cite{D0-Zg} using $Z \to ee,\mu\mu$.
Errors common to both analyses (e.g., luminosity, pdf uncertainties)
were taken into account when combining the results. 
The combined 95\% CL limits
are about $10\%$ tighter than for the neutrino channel alone and are listed
in Table~\ref{table2}.

Finally,   the   sensitivity  of  this   measurement to  the  value  of the
form-factor   scale  $\Lambda$ was  studied.  The  value  $\Lambda =
500$~GeV chosen above, 
is close  to the   sensitivity  limit  of  the  previous  Tevatron
measurements~\cite{CDF-Zg,D0-Zg}.
The  sensitivity of  the present measurement is higher and reaches  $\Lambda =
750$~GeV  for  the  neutrino  channel alone (slightly  higher for the
combined $ee+\mu\mu+\nu\nu$ channels). The 95\% CL limits obtained for
$\Lambda = 750$~GeV are  much tighter (see Table~\ref{table2}) and
are shown in Fig.~\ref{fig2}b.

It is important  to extend the  experimental  sensitivity to high values of
the  form-factor scale  which is  closely related  to the  scale of the new
physics which can  produce anomalous  couplings. Our  results show that the
sensitivity of  direct  measurements of  $Z\gamma$  production to anomalous
couplings   grows  with   $\Lambda$.  This  fact  makes such   measurements
complementary  to the  direct  searches for new  physics  which have higher
sensitivity   at low  scales.  The  limits  on  $h^V_{40}$  and  $h^V_{20}$
couplings for $\Lambda = 750$~GeV  obtained in this measurement are already
close to   expectations for   anomalous  couplings from  new  physics (see,
e.g.~\cite{Chang})    and  are  the  most  stringent  limits  on  anomalous
$ZV\gamma$ couplings currently available.

We thank U.~Baur and  J.~Ohnemus for Monte
Carlo  programs  and helpful  discussions.  We   thank the  staffs at
Fermilab and  collaborating  institutions  for their  contributions to this
work, and  acknowledge support from  the Department of  Energy and National
Science Foundation (U.S.A.), Commissariat \` a L'Energie Atomique (France),
State Committee for  Science and Technology and  Ministry for Atomic Energy
(Russia),  CNPq  (Brazil),  Departments  of Atomic  Energy and  Science and
Education  (India), Colciencias  (Colombia), CONACyT  (Mexico), Ministry of
Education and KOSEF  (Korea), CONICET and UBACyT  (Argentina), and the A.P.
Sloan Foundation.

\begin{table}[h]
\vspace*{-0.1in}
\caption{Summary of signal and backgrounds.}
\label{table1}
\begin{tabular}{lccc}
 & CC & EC & Total \\
\tableline
Candidates & 3 & 1 & 4 \\
\tableline
Muon background& $1.4 \pm 0.6$ & $0.4 \pm 0.2$ & $1.8 \pm 0.6$ \\
$W\to e\nu$ background& $2.2\pm 0.6$ & $1.8\pm 0.6$ & $4.0 \pm 0.8$ \\
$jj+j\gamma$ background & $< 0.4$ & $<0.2$ & $< 0.6$\\
\tableline
Total background:        & $3.6 \pm 0.8 $ & $2.2 \pm 0.6$ & $5.8 \pm 1.0$ \\
\tableline
SM signal prediction  & $1.4 \pm 0.2$  & $0.4 \pm 0.1$ & $1.8 \pm 0.2$\\
\end{tabular}
\end{table}

\begin{table}[h]
\vspace*{-0.25in}
\caption{95\% CL axis limits on the {\it CP\/}-conserving anomalous
couplings $h^V_{30}$, $h^V_{40}$. Limits on the {\it CP\/}-violating
partners $h^V_{10}$, $h^V_{20}$ are numerically the same.}
\label{table2}
\begin{tabular}{l|c|c|c|c}
Channel & $h^Z_{40} = 0$ & $h^Z_{30} = 0$ & $h^\gamma_{40}=0$ & $h^\gamma_{30}=0$\\
\tableline
\multicolumn{5}{c}{$\Lambda = 500$~GeV}\\
\tableline
$\nu\nu$ & $|h^Z_{30}| < 0.87$ & $|h^Z_{40}| < 0.21$ & 
$|h^\gamma_{30}| < 0.90$ & $|h^\gamma_{40}| < 0.22$\\
$ee,\mu\mu,\nu\nu$ & $|h^Z_{30}| < 0.78$ & $|h^Z_{40}| < 0.19$ & 
$|h^\gamma_{30}| < 0.81$ & $|h^\gamma_{40}| < 0.20$\\
\tableline
\multicolumn{5}{c}{$\Lambda = 750$~GeV}\\
\tableline
$\nu\nu$ & $|h^Z_{30}| < 0.49$ & $|h^Z_{40}| < 0.07$ & 
$|h^\gamma_{30}| < 0.50$ & $|h^\gamma_{40}| < 0.07$\\
$ee,\mu\mu,\nu\nu$ & $|h^Z_{30}| < 0.44$ & $|h^Z_{40}| < 0.06$ & 
$|h^\gamma_{30}| < 0.45$ & $|h^\gamma_{40}| < 0.06$\\
\end{tabular}
\end{table}

\begin{figure}[h]
\vspace*{3.5in}
\includegraphics{prl_pt.eps}
\caption{Transverse energy spectrum of photons in the $\MEts\gamma$ events.
The points show the data; the hatched curve 
is the SM signal prediction; the solid  line is the sum of
the SM signal prediction and the background,  with  the  errors shown
by  the band.  The inset  shows the predicted   $d\sigma/dE_T^\gamma$ 
folded  with the efficiencies for SM and anomalous couplings.}
\label{fig1}
\end{figure}

\begin{figure}[h]
\vspace*{3.5in}
\includegraphics{prl_limits.eps}
\caption{Limits    on the   correlated  ${\it     CP}$-con\-ser\-ving
ano\-ma\-lous $ZZ\gamma$ coupling parameters $h^Z_{30}$
and $h^Z_{40}$ for (a) $Z(\nu\nu)\gamma$ ($\Lambda = 500$~GeV) and
(b) $Z(ee+\mu\mu+\nu\nu)\gamma$ ($\Lambda = 750$~GeV). 
The solid  ellipses represent 95\% CL one- and
two-DOF exclusion contours.  The thin lines show
unitarity bounds.}
\label{fig2}
\end{figure}

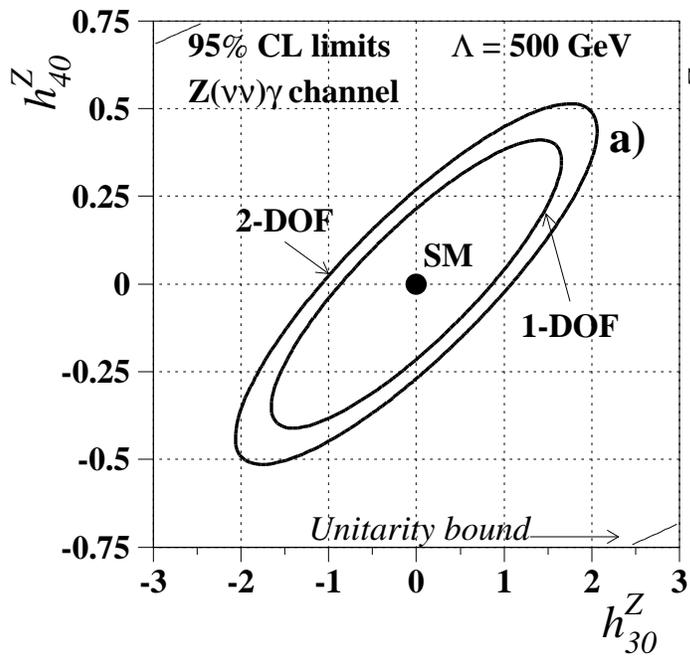
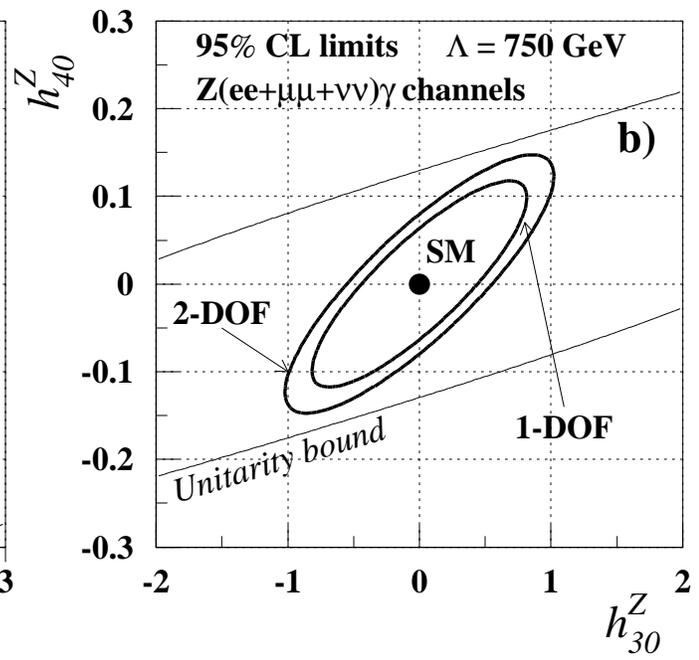
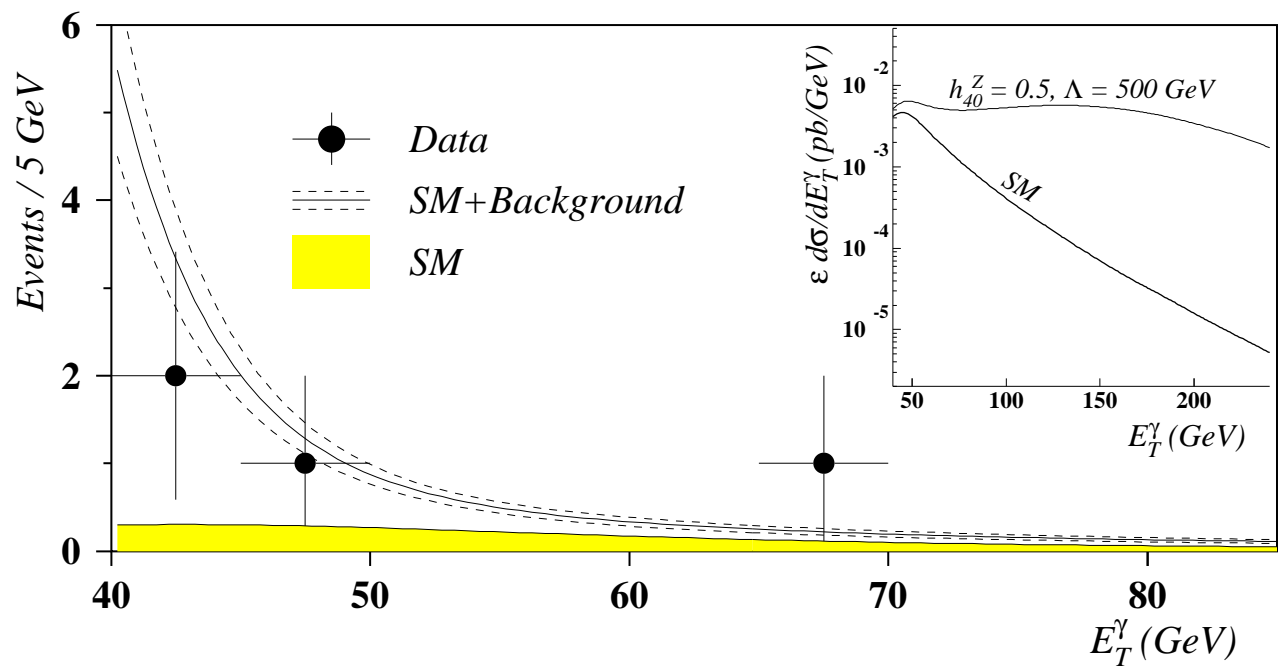
\end{document}